\documentclass[english,final]{aipproc}
\usepackage[T1]{fontenc}
\usepackage[latin9]{inputenc}
\usepackage{graphicx}
\usepackage{amssymb}

\makeatletter

\makeatletter

\usepackage{mathrsfs}

\layoutstyle{6x9}

\makeatother

\makeatother

\usepackage{babel}

\begin{document}
\newcommand{\apj}{ApJ\ }
 \newcommand{\apjl}{ApJ\ }
 \newcommand{\apjs}{ApJS\ }
 \newcommand{\aap}{A\&A\ }
 \newcommand{\mnras}{MNRAS\ }
 \newcommand{\nat}{Nature\ }
 \newcommand{\physrep}{Physics Reports\ }
 \newcommand{\Mo}{M_{\odot}}
 \newcommand{\Ro}{R_{\odot}}
 \newcommand{\Lo}{L_{\odot}}
 \newcommand{\Mbh}{M_{\bullet}}
 \newcommand{\SgrA}{SgrA$^{\star}$}
\newcommand{\Ms}{M_{\star}}
\newcommand{\Rs}{R_{\star}}
\newcommand{\Ls}{L_{\star}}
\newcommand{\Ns}{N_{\star}}
\newcommand{\nstr}{n_{\star}}

\title{The Galactic Center as a laboratory for extreme mass ratio gravitational wave source dynamics}
\author{Tal Alexander}{address={Weizmann Institute of Science, Faculty of Physics}, altaddress={The  William Z. and Eda Bess Novick Career Development Chair}} 

\begin{abstract}
The massive Galactic black hole and the stars around it are a unique
laboratory for studying how relaxation processes lead to close interactions
of stars and compact remnants with the central massive black hole,
in particular those leading to the emission of gravitational waves.
I review new results on the processes of strong mass segregation and
loss-cone refilling by massive perturbers and resonant relaxation;
describe observational evidence that these processes play a role in
the Galactic Center and can be studied there; and discuss some of
the implications for Extreme Mass Ration Inspiral event rates and
their properties. 
\end{abstract}
\maketitle

\section{introduction}

The $\Mbh\sim4\times10^{6}\,\Mo$ massive black hole (MBH) in the
Galactic Center (GC) \cite{eis+05,ghe+05} is special not only because
it is the nearest and observationally most accessible of all MBHs,
but also because it is an archetype of the extragalactic MBH targets
for the planned Laser Interferometer Space Antenna (LISA) gravitational
wave (GW) detector. This is a coincidence: current technology limits
LISA's baseline to $5\times10^{6}\,\mathrm{km}$, which happens to
coincide with the typical GW wavelength emitted by an object on the
last stable circular orbit around a $\sim10^{7}\, M_{\odot}$ MBH.
GW from more massive MBH have longer wavelengths, beyond the sensitivity
of LISA. As shown below, LISA's focus on the lower range of MBH masses
implies that LISA targets will typically lie in dynamically relaxed,
high density nuclei. In such systems, relaxation controls the rate
at which compact remnants on wide orbits near the MBH are scattered
into inspiral orbits ({}``loss-cone'' orbits) and become extreme
mass ratio inspiral (EMRI) GW sources. Thus, to understand and predict
EMRI rates and orbital properties, it is necessary to understand the
various dynamical relaxational processes that occur close to a MBH,
and the degree of concentration of compact remnants around it. I describe
here several newly (re-)discovered relaxation mechanisms beyond standard
2-body relaxation, which likely play an important role in EMRI dynamics,
and discuss how these theoretical ideas are tested by observations
of the GC.

\section{Nuclear dynamics of LISA MBH targets}

\label{s:lowM}

LISA targets have dynamically relaxed, high density stellar cusps.
This is a direct consequence of the tight empirical correlation observed
between MBH masses and the velocity dispersion of the spheroid of
their host galaxy, $\Mbh\propto\sigma^{\beta}$ ($4\!\lesssim\!\beta\!\lesssim5$)
\cite{fer+00,geb+00}. Here we assume for simplicity $\beta\!=\!4$;
a higher value only reinforces the following conclusions.

The MBH radius of dynamical influence is defined as $r_{h}\!\sim\! GM_{\bullet}/\sigma^{2}\!\propto\!\Mbh^{1/2}$.
The mass in stars within $r_{h}$ is of the order of the mass of the
MBH, so their number is $N_{h}\!\sim\! M_{\bullet}/\Ms$, where $\Ms$
is the mean stellar mass, and the average stellar density there is
$\bar{n}_{h}\!\sim\! N_{h}/r_{h}^{3}\!\propto\! M_{\bullet}^{-1/2}$.
The {}``$nv\Sigma$'' rate of strong gravitational collisions can
be used to estimate the two-body relaxation time at $r_{h}$, $T_{R}\!\sim\!\left[\bar{n}_{h}\sigma(G\Ms/\sigma^{2})^{2}\right]^{-1}\propto\!\Mbh^{5/4}$
(note for future reference that $T_{R}^{-1}\!\propto\Ms^{2}\Ns$).
Evaluated for the Galactic MBH, $T_{R}\!\sim\! O(1\,\mathrm{Gyr})\!<t_{H}$
(the Hubble time) and $\bar{n}_{h}\!\sim\! O(10^{5}\,\mathrm{pc^{-3}})$.
Note that for a MBH only a few times more massive than the Galactic
MBH, $T_{R}\!\gtrsim\! t_{H}$. The GC is thus a member of a subset
of galactic nuclei with relaxed, high-density stellar cusps. Such
systems are expected to settle into a power-law cusp distribution,
$n_{\star}\!\propto\! r^{-\alpha}$ \cite{bah+77}. Therefore, close
to the MBH where EMRIs originate, $\nstr\!\gg\!\bar{n}_{h}$. The
short relaxation time also implies that the system will not retain
memory of any past major perturbations, such as a merger with a second
MBH \cite{mer+06,mer+07}.

\section{Strong mass segregation}

\label{s:MassSeg}

\begin{figure}
\includegraphics[width=0.5\textwidth]{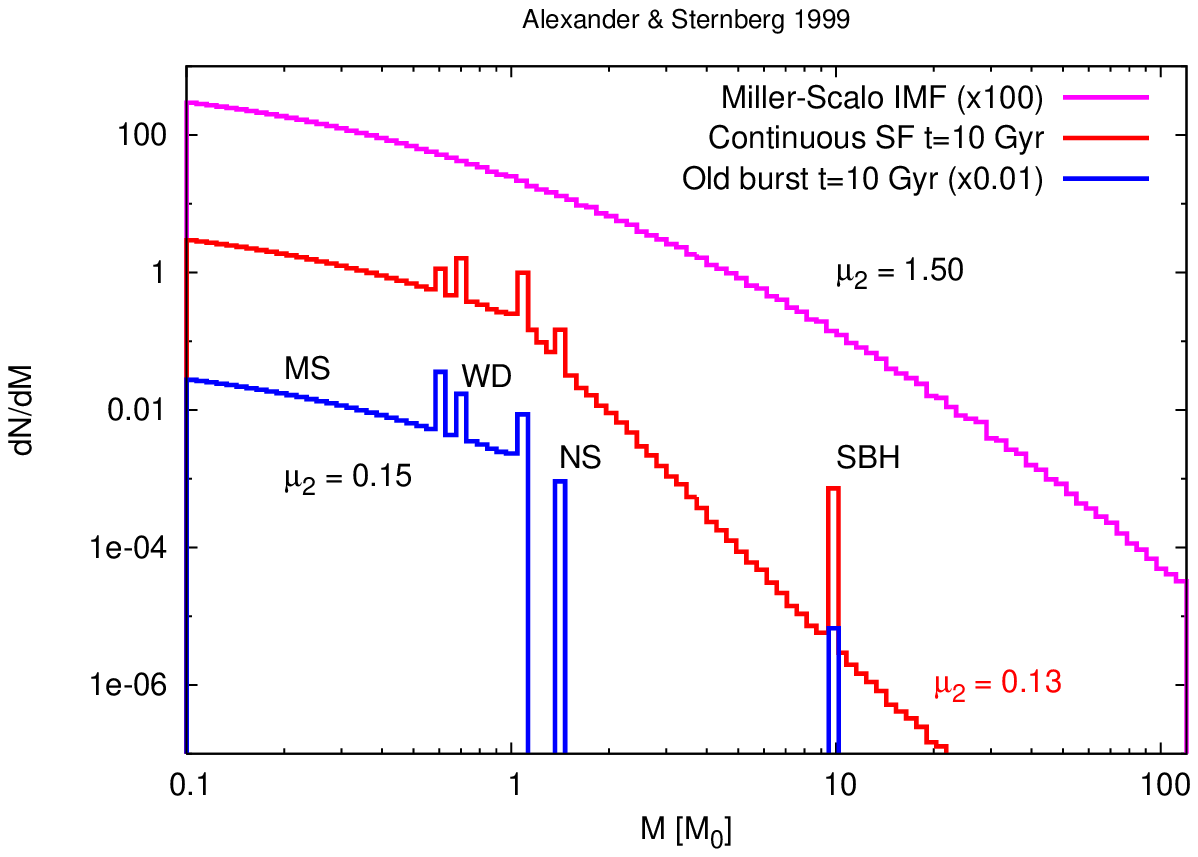} \includegraphics[width=0.5\textwidth]{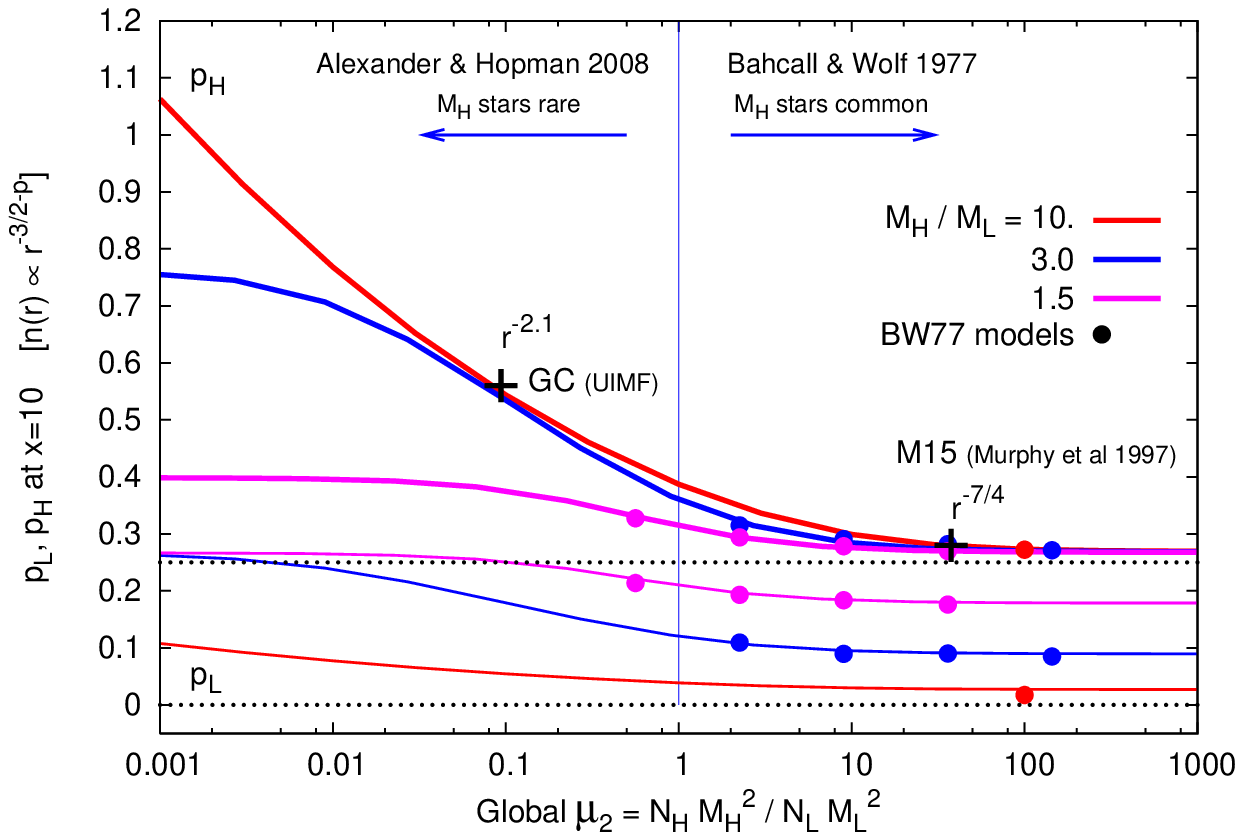} 

\caption{\label{f:MassSeg} \textbf{Left}: Strong mass segregation. The predicted
values of the global relaxational self-coupling parameter $\mu_{2}$
for a {}``universal'' Miller \& Scalo IMF \cite{mil+79} (top line),
an evolved mass function assuming continuous stars formation over
10 Gyr (middle line), and an evolved star formation burst 10 Gyr old
(bottom line) {[}Alexander \& Hopman 2008, in prep.]. The mass functions
of the old populations develop excesses in the $\sim0.6$--$1.4\, M_{\odot}$
range due to the accumulation of white dwarfs and neutron stars, and
in the $\sim10M_{\odot}$ range due to the accumulation of stellar
black holes (here represented by a simplified discrete mass spectrum,
see \cite[tbl 2.1]{ale05}). \textbf{Right}: Fokker-Planck mass-segregation
results {[}Alexander \& Hopman 2008, in prep.]. The logarithmic slopes
$p_{H}$ and $p_{L}$ of the distribution functions of the heavy stars
(thick lines) and light stars (narrow lines) , evaluated at ($r\sim0.1$
pc in the GC), as function of the global relaxational self-coupling
parameter $\mu_{2}$, for mass ratios of $M_{H}/M_{L}$ = 1.5, 3,
10. The cross on the left indicates the logarithmic slope of the stellar
density of massive stars in the GC ($\alpha_{H}=3/2+p_{H}\simeq2.2$
for $\mu_{2}\sim0.13$, see left panel), assuming a universal IMF
and continuous star formation history. The cross on the right indicates
the logarithmic slope for globular cluster M15, assuming it harbors
an IMBH ($\alpha_{H}=1.75$ estimated at $\mu_{2}\sim37$, based on
a model for the cluster mass function \cite{mur+97}). The results
for the models studies by BW76 \cite{bah+77} are indicated by circles. }

\end{figure}

The degree of central concentration of compact remnants, and in particular
of stellar mass BHs (SBHs), is crucial for determining the EMRI rate,
since the analysis of the loss-cone refilling problem for inspiral
orbits \cite{hop+05} shows that the rate is $\Gamma\!\sim\!\Ns(<r_{\mathrm{crit}})/T_{R}$,
where $\Ns$ is the number of enclosed stars within $r_{\mathrm{crit}}\sim O(0.01\,\mathrm{pc})$,
the  critical radius that demarcates the boundary between SBHs that
plunge (infall) directly into the MBH due to perturbations by other
stars ($r\!>\! r_{\mathrm{crit}}$), and SBHs that gradually inspiral
unperturbed into the MBH due to the dissipation of orbital energy
by the emission of GW ($r\!<\! r_{\mathrm{crit}}$). A relaxed system
is expected to undergo mass-segregation, where the massive, long-lived
objects (SBHs) sink to the center, while the relatively light, long-lived
objects (main-sequence dwarfs, white dwarfs and neutron stars) are
pushed out. 

The Bahcall-Wolf 1977 (BW77) solution \cite{bah+77}, long-assumed
to universally describe relaxed cusps around a MBH, predicts that
stars in a population with masses $M_{L}\!\le\!\Ms\!\le\! M_{H}$
should relax to power-law cusps with a mass-dependent slope, $n_{M}\!\propto\! r^{-\alpha_{M}}$,
where $\alpha_{M}=3/2+p_{M}$ and $p_{M}\!=\!\Ms/4M_{H}$. Thus, SBHs
are predicted to have a steep $r^{-7/4}$ cusp, while the lowest-mass
stars will have a flatter $r^{-3/2}$ cusp. In the single mass limit,
this solution agrees with the known result $\alpha\!=\!7/4$ \cite{bah+76}.
However, the BW77 solution is both puzzling and problematic since
(i) it does not depend at all on the mass function (apparently, even
a few very massive objects can flatten the cusp of the dominant low-mass
stars), and (ii) it predicts a lower degree of segregation than the
strong segregation ($\alpha_{10\Mo}\!>\!2$) that is measured in simulations
\cite{fre+06} and analytical models \cite{hop+06b} of the GC. Such
strong segregation is also is consistent with the observed strong
central suppression of low-mass giants in the GC \cite[ M. Levi, MSc thesis]{sch+07}.

A recent study of the mass segregation problem \cite[Alexander \& Hopman 2008, in prep.]{ale07}
reveals a new branch of solutions which apply in the limit where the
heavy objects are relatively rare, not considered by BW76. This the
generic case in old coeval or continuously star-forming populations
with universal IMFs, such as observed in the GC \cite{ale+99a}. In
old populations, the initially broad mass function {}``polarizes''
over time into two mass groups: $O(1\,\Mo)$ objects (main sequence
dwarfs, white dwarfs and neutron stars) and $O(10\,\Mo$) objects
(SBHs) (Fig. \ref{f:MassSeg} left). When the heavy objects are rare,
$M_{H}$--$M_{L}$ interactions dominate, leading to dynamical friction-induced
sinking of the heavy stars to the center, and to the formation of
a steep cusp. The relative strength of $M_{H}$--$M_{H}$ scattering
and $M_{H}$--$M_{L}$ drag can be roughly expressed by a {}``relaxational
self coupling parameter'', $\mu_{2}\!\equiv\! M_{H}^{2}N_{H}/M_{L}^{2}N_{L}$
, where $N_{H}/N_{L}$ is the number ratio of heavy to light stars
far from the MBH, where mass segregation is negligible (cf the derivation
of the relaxation time above). BW77 explored only models with $\mu_{2}\gtrsim1$,
where the heavy stars dominate the dynamics and behave essentially
as a single mass population. However, old populations have $\mu_{2}\!\sim\! O(0.1)$
(Fig. \ref{f:MassSeg} left). This leads to a much steeper mass segregation
slope (Fig. \ref{f:MassSeg} right), which is consistent with what
is found in simulations and observations. If relaxed LISA galactic
nuclei targets are similar to the GC, then the expected degree of
central concentration within $r_{\mathrm{crit}}$ will be higher than
previously anticipated. The implications of this for EMRI rates require
further study.

\section{Massive perturbers}

\label{s:MPs}

\begin{figure}
\includegraphics[width=0.5\textwidth]{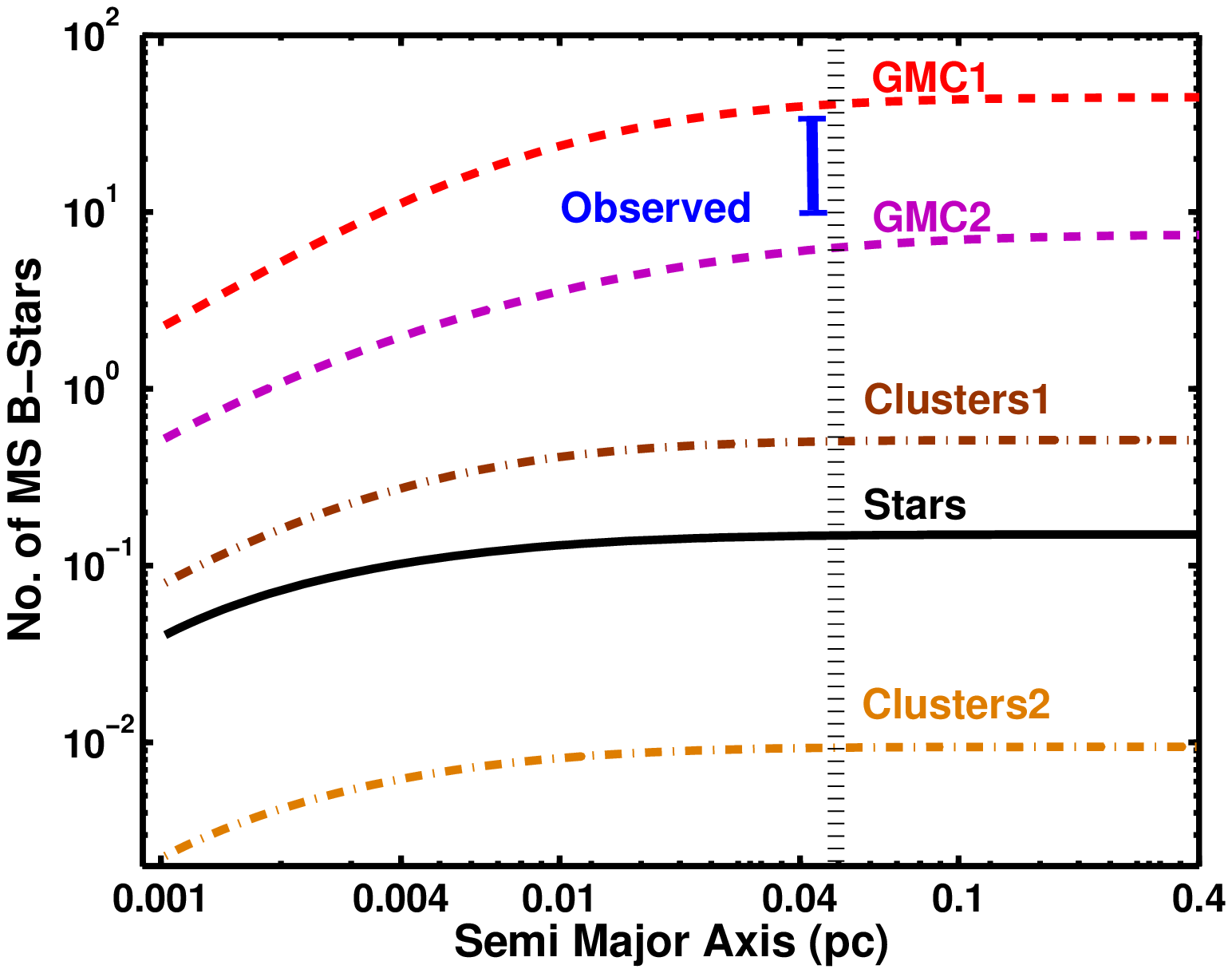} \includegraphics[width=0.5\textwidth]{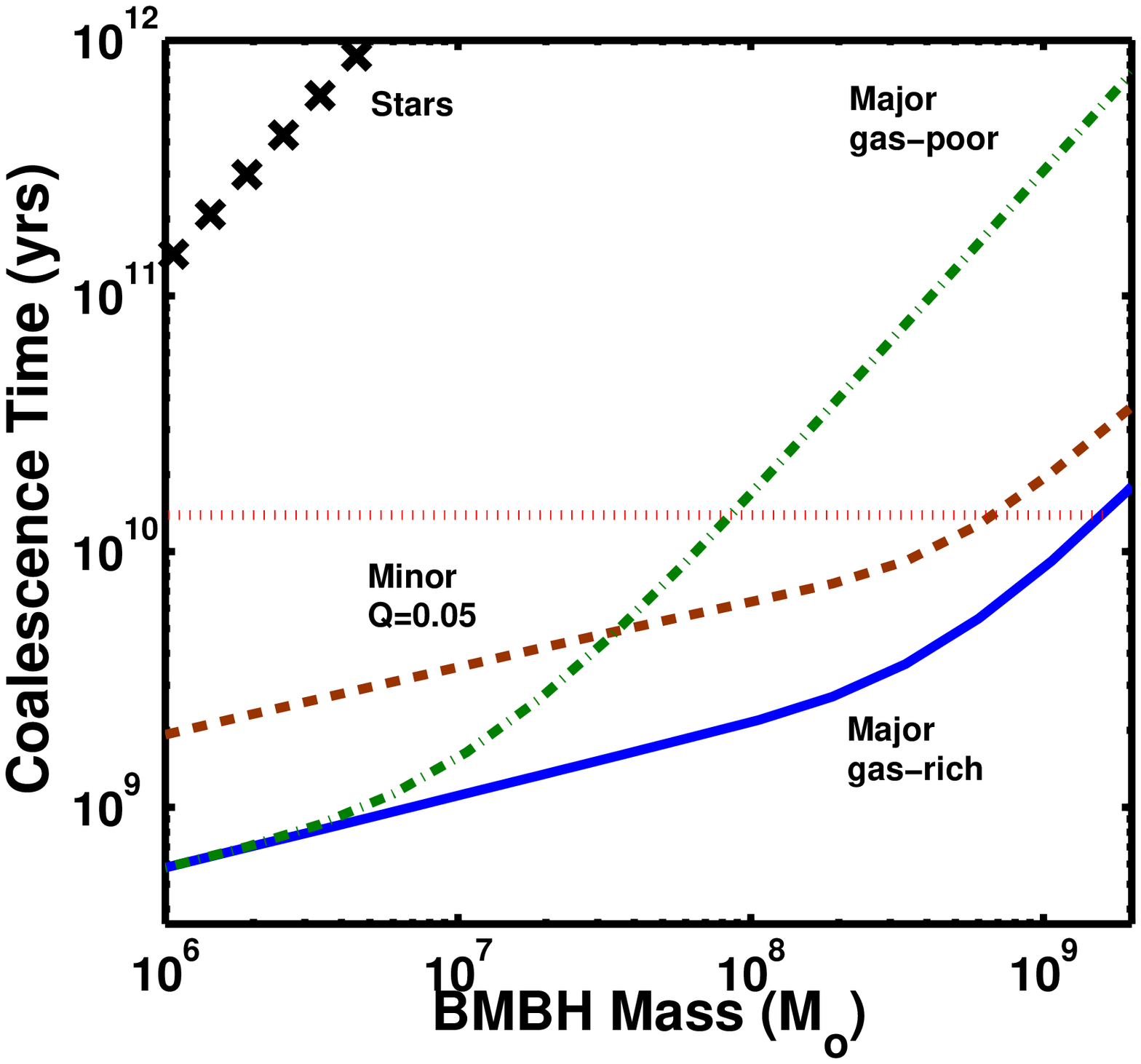} 

\caption{%
 \label{f:MPs} \textbf{Left}: A comparison between the cumulative
number of S-stars (main sequence B stars) observed orbiting the Galactic
MBH on randomly oriented orbits (vertical bar), and the predicted
number captured by 3-body tidal interactions of the MBH with binaries
detected to the center by massive perturbers, for different massive
perturbers models \cite{per+07}. The observed extent of the S-star
cluster is indicated by the vertical hashed line. \textbf{Right}:
Accelerated binary MBH mergers in the presence of MPs \cite{per+08}.
The time to coalescence as function of binary MBH mass, for different
merger scenarios distinguished by the mass ratio $Q$ between the
two MBHs and the MP contents of host galaxies. The age of the universe
is indicated by the dotted horizontal line. Stellar relaxation alone
cannot supply a high enough rate of stars for the slingshot mechanism
to complete the merger within a Hubble time. However, in minor mergers
($Q=0.05$) and major gas-rich mergers ($Q=1$) with MPs, merger is
possible within a Hubble time for all but the most massive MBHs. (Reproduced
with permission from the \emph{Astrophysical Journal})}

\end{figure}

The $T_{R}^{-1}\!\propto M^{2}N$ scaling of the relaxation rate implies
that a few very massive perturbers (MPs) of mass $M_{p}\!\gg\!\Ms$,
such as clusters, giant molecular clouds (GMCs), or intermediate mass
BHs (if those exist), may well dominate the relaxation and fundamentally
affect the loss-cone refilling rate, if $\mu_{2}\!=\! M_{P}^{2}N_{P}/\Ms^{2}\Ns\!\gg\!1$.
For example, there are $O(100)$ GMCs with masses $ $$10^{4}\!\lesssim\! M_{p}\!\lesssim\!10^{7}\,\Mo$
observed in the central $\sim\!100$ pc of the Galaxy \cite{oka+01},
as compared to $\Ns\!\sim\!10^{8}$ solar mass stars, so $100\!\lesssim\!\mu_{2}\!\lesssim\!10^{8}$
on that scale. 

We show \cite{per+07} that MPs, which affect the dynamics on the
$r_{\mathrm{MP}}\gtrsim\mathrm{few\, pc}$ scale (GMCs are tidally
disrupted closer in), strongly increase the rate of close interactions
with loss-cone processes for which $r_{\mathrm{crit}}\!>\! r_{\mathrm{MP}}$.
MPs are thus not relevant for EMRIs, which are scattered singly into
inspiral orbits from $r_{\mathrm{crit}}\sim O(0.01\,\mathrm{pc})$,
or for tidally disrupted stars, which originate from $\sim r_{h}<r_{\mathrm{MP}}$
\cite{lig+77}. However, the rates of close interactions that occur
at larger periapse and so have a larger $r_{\mathrm{crit}}$, such
as the tidal separation of binary by the MBH, are strongly enhanced
by MPs. The tidal separation of a binary with mass $M_{2}$ and semi-major
axis $a_{2}$ by an MBH of mass $\Mbh$ results in the capture of
one star on a tight and eccentric orbit with semi-major axis $\left\langle a\right\rangle \!\sim\!0.6(\Mbh/M_{2})^{2/3}a_{2}$,
and eccentricity $\left\langle e\right\rangle \sim1-2(M_{2}/\Mbh)^{1/3}$,
and the ejection of the other star at an extremely high velocity,
$v_{\infty}^{2}\!\sim\!\sqrt{2}GM_{2}^{2/3}\Mbh^{1/3}/a_{2}$, which
exceeds the escape velocity from the Galaxy \cite{hil88}. MP-induced
scattering of massive binaries from the field toward the Galactic
MBH and their ensuing tidal separation can naturally explain \cite{per+07}
the number and spatial extent of the mysterious cluster of young stars
observed within $\mathrm{few}\times0.01$ pc of the Galactic MBH \cite{eis+05}
(Fig. \ref{f:MPs}, left), as well as the number of hyper-velocity
stars observed at distances of tens of kpc from the GC, on their way
out of the Galaxy \cite{bro+05}. 

If typical LISA targets are similar to the Galaxy, then binaries containing
compact object, which are scattered to the MBH by MPs and tidally
separated, will increase the EMRI rate from WDs by an order of magnitude
relative to that of singly scattered EMRIs {[}Perets, Hopman \& Alexander,
2008, in prep.].

The orbital decay of a binary MBH in a post-merger galaxy by 3-body
interactions with stars scattered into its orbit ({}``the slingshot
effect'') is another large loss-cone process that is accelerated
by MPs. Binary MBH coalescence events are predicted to be the strongest
sources of GW in the universe, observable by LISA from extremely high
redshifts, tracing the galactic merger history and growth history
of MBHs. Dynamical simulations indicate that binary MBHs in spherically
symmetric systems with no gas on small scales ({}``dry mergers'')
stall, and do not reach within a Hubble time the final rapid phase
of GW dissipation leading to coalescence \cite{mer+05b}. This situation,
known as the {}``last pc problem'' is circumvented by MPs on large
scales, which can efficiently scatter stars into the binary's orbit,
in all but the most massive galaxies \cite{per+08} (Fig. \ref{f:MPs},
right).

\section{Resonant relaxation}

\label{s:RR}

\begin{figure}
\includegraphics[width=0.5\textwidth]{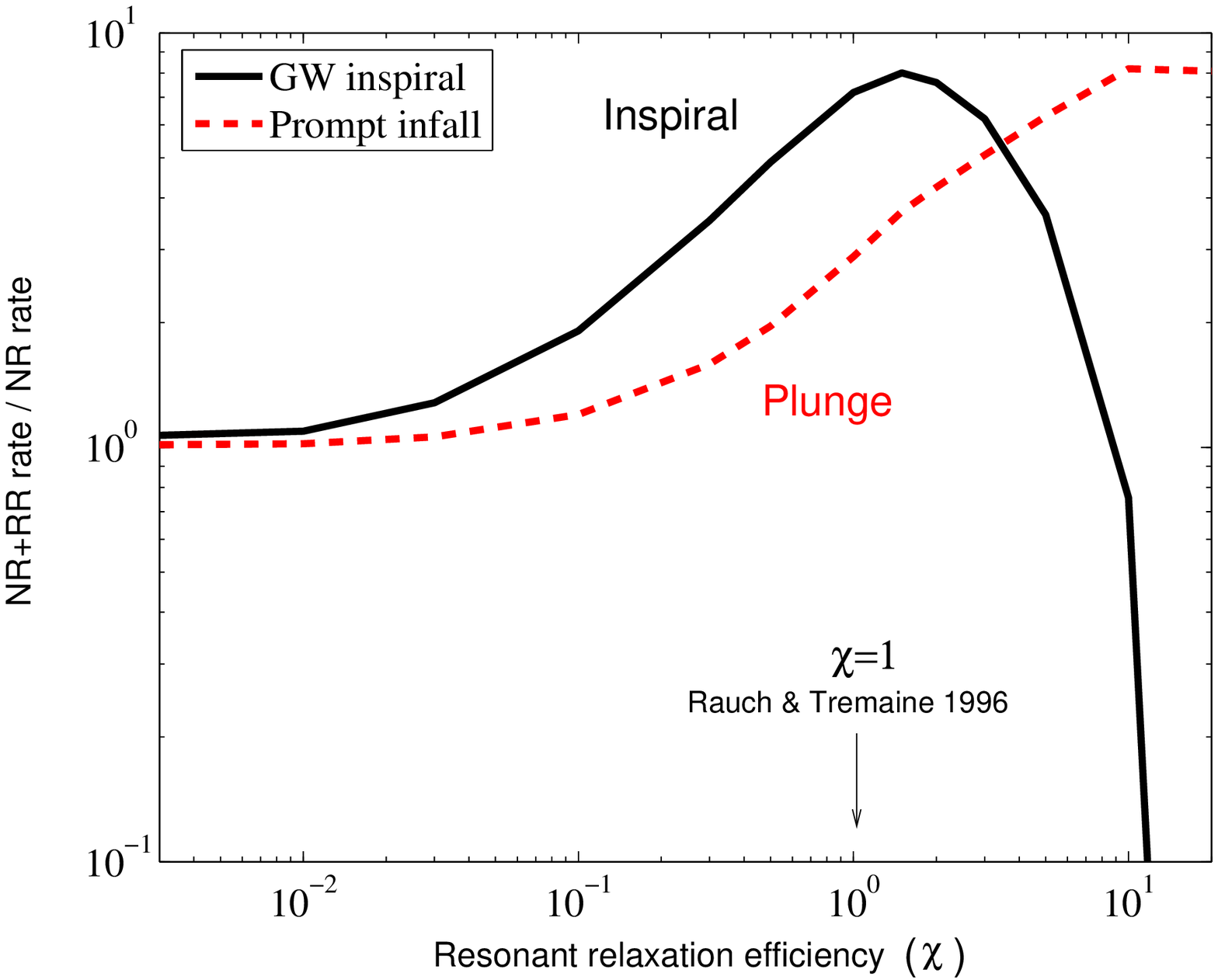} \includegraphics[width=0.5\textwidth]{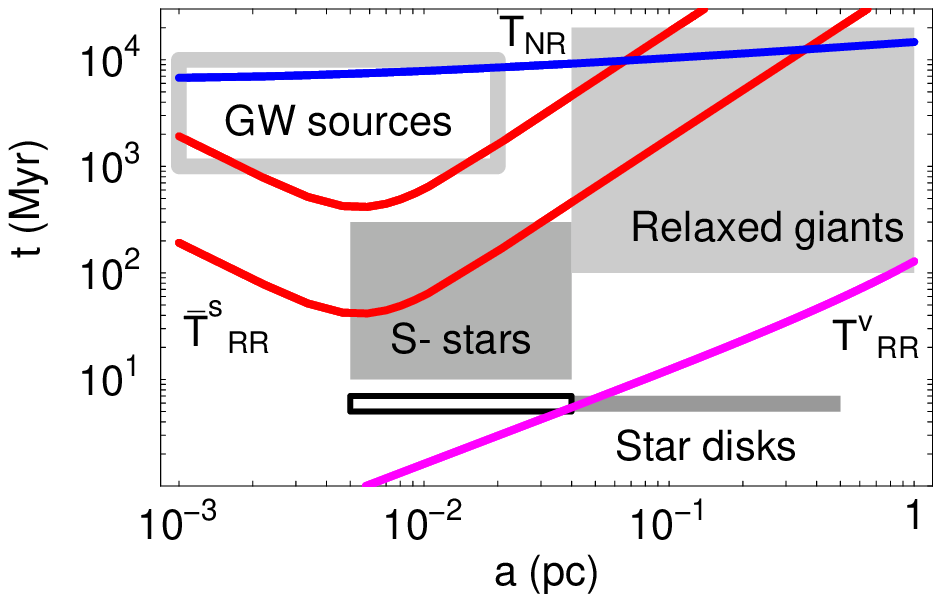} 

\caption{%
 \label{f:RR} \textbf{Left}: Resonant relaxation efficiency. The
relative rates of GW EMRI events and direct infall (plunge) events,
as function of the unknown efficiency of RR, $\chi$, normalized to
$\chi=1$ for the values derived by Rauch \& Tremaine \cite{rau+96}.
\textbf{Right}: Evidence for resonant relaxation in the GC in the
plane of age .vs. distance from the MBH. The spatial extent and estimated
age of the various dynamical sub-populations in the GC (shaded areas)
is compared with the NR timescale (top line, for assumed mean mass
of $M=1\, M_{\odot}$) and with the scalar RR timescale (two curved
lines, top one for $\chi M_{\star}=1\, M_{\odot}$, bottom one for
$\chi M_{\star}=10M_{\odot}$ ) and vector RR timescale (bottom line,
for $\chi M_{\star}=1\, M_{\odot}$). The populations include the
young stellar rings in the GC (filled rectangle in the bottom right);
the S-stars, if they were born with the disks (open rectangle in the
bottom left); the maximal lifespan of the S-stars (filled rectangle
in the middle left); the dynamically relaxed red giants (filled rectangle
in the top right); and the reservoir of GW inspiral sources, where
the age is roughly estimated by the progenitor's age or the time to
sink to the center (open rectangle in the top left). Stellar components
that are older than the various relaxation times must be randomized
\cite{hop+06a}. (Reproduced with permission from the \emph{Astrophysical
Journal})}

\end{figure}

Two-body relaxation, or non-coherent relaxation (NR), is inherent
to any discrete large-N system, due to the cumulative effect of uncorrelated
two-body encounters. These cause the orbital energy $E$ and the angular
momentum $J$ to change in a random-walk fashion ($\propto\!\sqrt{t}$)
on the typically long NR timescale $T_{\mathrm{NR}}(r)\!\simeq\!(\Mbh/\Ms)^{2}P/\Ns\log\Ns$,
where $P$ is the orbital time at radius $r$ and $\Ns$ the number
of stars enclosed there. In contrast, when the gravitational potential
has approximate symmetries that restrict orbital evolution (e.g. fixed
ellipses in a Keplerian potential; fixed orbital planes in a spherical
potential), the perturbations on a test star are no longer random,
but correlated, leading to coherent ($\propto\! t$) torquing of $J$
on short timescales, while the symmetries hold. Over longer times,
this results in resonant relaxation (RR) \cite{rau+96,rau+98}, a
rapid random walk of $J$ on the typically short RR timescale $T_{\mathrm{RR}}\!\ll\! T_{\mathrm{NR}}$. 

The properties and timescales of RR depend on the symmetries of the
potential and the processes that break them. RR in a near-Keplerian
potential can change both the direction and magnitude of $\mathbf{J}$
({}``scalar RR''), thereby driving stars to near-radial orbits that
interact strongly with the MBH. Far from the MBH, coherent torquing
is limited by orbital precession due to the enclosed stellar mass,
and the resulting scalar RR timescale is $T_{\mathrm{RR}}^{J}\!\simeq\!(\Mbh/\Ms)P$.
Close to the MBH, where General Relativistic (GR) precession limits
the coherent torquing, $T_{\mathrm{RR}}^{J}\!\simeq\!(3/8)(\Mbh/\Ms)^{2}(J_{\mathrm{LSO}}/J)^{2}P/\Ns$,
where $J_{\mathrm{LSO}}\!=\!4G\Mbh/c$ is the angular momentum of
the last stable orbit. The combined effects of mass and GR precession
result in a minimal RR timescale, $\min T_{\mathrm{RR}}^{J}\!\ll\! T_{\mathrm{NR}}$,
at $r\sim O(0.01)$ pc for a typical LISA target, which happens to
coincide with $r_{\mathrm{crit}}$ for inspiral \cite{hop+06a} (Fig.
\ref{f:RR}, right). Thus, it is scalar RR, and not NR, that dominates
EMRI dynamics. RR in a near-spherical potential (as is most likely
the case anywhere within $r_{h}$) can only change the direction of
$\mathbf{J}$ ({}``vector RR''). Coherent torquing is limited by
vector RR itself, which randomizes the orbital planes (alternatively,
this can be viewed as the result of potential fluctuations due to
the finite number of stars), and results in $T_{\mathrm{RR}}^{\mathbf{J}}\!\simeq\!2(\Mbh/\Ms)P/\sqrt{\Ns}$.

There are as yet only a few quantitative studies of RR. While the
effect is clearly seen in $N$-body simulations \cite{rau+96,eil+08},
systematic studies of RR efficiency and its dependence on the properties
of the stellar cusp around the MBH, and on the orbital parameter of
the torqued star, are only now yielding first results \cite{gur+07,eil+08}.
Determining the exact value of RR efficiency (parametrized by a dimensionless
order unity coefficient $\chi$, Fig. \ref{f:RR}) is crucial for
predicting its effects on EMRI rates. RR can increase the EMRI rate
by rapidly deflecting stars to near-radial inspiral orbits that lead
to quasi-periodic GW signals. However, if RR is too efficient, then
it strongly suppresses the EMRI rate from single compact objects by
throwing all compact objects within $r_{\mathcal{\mathrm{crit}}}$
directly into the MBH, producing instead short, hard to detect GW
bursts \cite{hop+06a} (Fig. \ref{f:RR}, left). The effect of strong
RR on EMRIs from tidally separated binaries (see above) has yet to
be investigated in detail. Recent numerical studies indicate an RR
efficiency of $\chi\!\simeq\!4$, which corresponds to an RR enhancement
of the EMRI rate by a factor of few over that predicted for NR only
\cite{eil+08}.

The fact that scalar RR is quenched by GR precession when $J\!\rightarrow\! J_{\mathrm{LSO}}$,
a limit that also coincides with the stage where GW dissipation becomes
important, is another crucial but still little-studied aspect of EMRI
dynamics. This is important because GR quenching allows compact remnants
to be deflected very rapidly by RR to strongly relativistic orbits,
but then stall \textquoteleft{}\textquoteleft{}on the brink\textquoteright{}\textquoteright{}
and instead of falling directly into the MBH, inspiral into it gradually
as EMRIs.

The systematic dynamical differences between the various stellar components
observed in the GC (the young central S-star cluster in the inner
$\mathrm{few}\times0.01$ pc, the disk(s) of very massive young stars
on the $\sim0.0$5--0.5 pc scale, and the $\lesssim\mathrm{few}\times10^{9}$
yr old population of isotropic red giants) can be explained by the
effects of RR (Fig. \ref{f:RR}, right). In particular, the truncation
of the star disk(s) at $\sim\!0.04$ pc is consistent with scalar
RR with $\chi\!\sim\! O(1)$, and the relaxed population of red giants
is consistent with the effects of vector RR. Thus, observations of
the GC can provide an empirical test of the RR concept and constrain
its efficiency.

\section{Summary}

\label{s:summary}

The empirical $\Mbh/\sigma$ relation implies that LISA extragalactic
targets, MBHs in the range $\Mbh\!\lesssim\!10^{7}\,\Mo$, lie in
very high density relaxed stellar cusps. The Galactic MBH happens
to be an archetype of LISA targets. In such systems, it is the slowest
relaxation process, non-coherent 2-body relaxation, or possibly other
more efficient processes, that determine the central concentration
of compact EMRI candidates, and the rate at which they are captured
in EMRI orbits. We described several new results on processes that
may strongly affect the cosmological EMRI rates, and can be tested
by observations of the GC: strong mass segregation, which drives heavy
SBHs into unusually high concentrations near the MBH, and may have
left an imprint in the strong observed central suppression of light
giants around the Galactic MBH; accelerated loss-cone refilling by
massive perturbers, which strongly affects the rate at which binaries
are tidally separated by the MBH, leaving compact objects very close
to the MBH. This mechanism very naturally explains the cluster of
young stars around the Galactic MBH and the numbers of hyper-velocity
stars observed on their way out of the Galaxy; and RR, which dominates
EMRI dynamics and likely enhances their rate. RR can explain some
of the systematic differences between the dynamical properties of
the various stellar populations observed in the GC, and conversely,
these observations can be used to test this process and constrain
its properties.

The implications of all these processes on the EMRI rates have still
to be fully mapped. However, it is already safe to say that GW astronomy
will tell us as much about fundamental questions in stellar dynamics,
as it will tell us about cosmology and GR. 

\begin{theacknowledgments} This work was supported by ISF grant 928/06,
Minerva grant 8563 and a New Faculty grant by Sir H. Djangoly, CBE,
of London, UK. \end{theacknowledgments}

\bibliographystyle{aipproc}

\end{document}